\documentstyle[epsfig]{aipproc}
\def\gsim{\lower.5ex\hbox{$\; \buildrel > \over \sim \;$}}
\def\lsim{\lower.5ex\hbox{$\; \buildrel < \over \sim \;$}}
\begin{document}
\title{Surveys and the Blazar Parameter Space}

\author{Eric S. Perlman}
\address{Joint Center for Astrophysics, University of Maryland, 1000 Hilltop
Circle, Baltimore, MD  21250, USA}

\author{Paolo Padovani, Hermine Landt}
\address{STScI, 3700 San Martin Drive, Baltimore, MD  21218, USA}

\author{John T. Stocke}
\address{Center for Astrophysics and Space Astronomy, University of Colorado,
Campus Box 389, Boulder, CO 80309, USA}

\author{Luigi Costamente}
\address{Osservatorio di Brera, via Bianchi 46, 23807 Merate (LC), 
Italy}

\author{Travis Rector}
\address{NOAO, P. O. Box 26851, Tucson, AZ 85721, USA}

\author{Paolo Giommi}
\address{SAX Science Data Center, Agenzia Spatiale Italiana, viale Regina
Margherita 202, I-00198 Roma, Italy}

\author{Jonathan F. Schachter}
\address{Smithsonian Astrophysical Observatory, 60 Garden Street, Cambridge,
MA  02138, USA}

\maketitle

\begin{abstract}

The rareness of blazars, combined with the previous history of  relatively
shallow, single-band surveys, has dramatically colored our perception of these
objects.  Despite a quarter-century of research, it is not at all clear whether
current samples can be combined to give us a relatively unbiased view of blazar
properties, or whether they present a view so heavily affected by biases
inherent in single-band surveys that a synthesis is impossible.  We will use the
coverage of X-ray/radio flux space for existing  surveys to assess their
biases. Only new, deeper blazar surveys approach the level needed in depth and
coverage of parameter space to give us a less biased view of blazars.  These
surveys have drastically increased our knowledge of blazars' properties.  We
will specifically review  the discovery of ``blue'' blazars, objects with broad
emission lines but broadband spectral characteristics similar to HBL BL Lac
objects.

\end{abstract}

\section{Myths and Facts about Blazar Surveys}

Every survey has its biases, whether imposed by its flux limits or selection
techniques - despite the best efforts of the scientists involved.  Proper 
analysis of the results of any survey requires one to understand the impact
of these biases on both the range of parameter
space to which the survey is sensitive, and also on the
broader scheme, including properties which do not form part of the survey
definition, but which nevertheless represent characteristics of the class
and/or important diagnostics.  

Historically, our knowledge of blazar properties began with
radio surveys, which tend to be dominated by radio-bright objects.   As we
discovered in the 1980s, such objects are the most
luminous of all AGN, as well as violently variable and radio core
dominated.  Indications that these properties were not typical of all blazars
did not come until the publication of the first large X-ray survey, the {\it
Einstein} EMSS (Stocke et al. 1991).  The EMSS BL Lacs are
considerably less luminous, less variable, less polarized, and less core
dominated than their radio-selected cousins (Perlman \&
Stocke 1993; Jannuzi, Elston \& Smith 1994; 
Kollgaard et al. 1996; Giommi et al. 1995; Rector et al. 2000).  They
also have synchrotron peaks in the UV/X-ray rather
than IR/optical (Giommi et al. 1995, Sambruna et al. 1996, Fossati et al.
1998).  Various explanations for these differences have been proposed  (see the
review by Urry \& Padovani 1995, and also Ghisellini et al. 1998 and
Georganopoulos \& Marscher 1998).  But it was not until the mid-1990s that
opinions began to come full circle.

Even if one assumes an unbiased identification process  (not always the case,
see e.g., Marcha \& Browne 1993, 1996; Perlman et al. 1996;  Rector et al.
1999) each survey's flux limits affect its sensitivity to parameter space.  In
his paper, Paolo Padovani presented two figures (Padovani 2000, Figures 1 and
2) which illustrate the sensitivity of various surveys to X-ray/radio parameter
space, using their flux limits. For completeness and for the sake of minimum
bias, one wants surveys that are: (1) as close as possible to the diagonal
lines defined by the inverse-Compton and synchrotron peak limits - which are
also the extrema for known HBLs and LBLs respectively; and (2) large enough
that all spectral shape classes are represented in statistically significant
numbers.  Yet even with surveys meeting these requirements, some biases remain,
since each flux limit imposes diagonal ``completeness contours'' on the
$(\alpha_{ox},\alpha_{ro})$ plane.  Thus even a survey which is near  an
extreme line in the $(f_x,f_r)$ plane does not have uniform sensitivity to all
regions of $(\alpha_{ox},\alpha_{ro})$ space.  Nevertheless, once samples
meeting these criteria are created, it will truly be possible to use their
contents to simulate the overall blazar content of the universe.

Figure 1 illustrates these biases for the case of BL Lacs.  Due to their small
sizes and high flux limits, each of the ``classical'' surveys (EMSS, 1 Jy and
Slew) was sensitive  only to a small diagonal swath of
$(\alpha_{ox},\alpha_{ro})$ parameter space. As a result, they presented an
almost completely disjoint picture of BL Lac properties (it is an instructive
exercise to plot the X-ray, radio and optical fluxes of the EMSS, 1 Jy and Slew
BL Lacs on three planes - we omit this here for lack of space).  As this
diagram shows, the new DXRBS survey fills in the gap very nicely, covering a
much wider range of the $(\alpha_{ox},\alpha_{ro})$ plane much more evenly, and
nicely revealing objects in the intermediate range. That intermediate range is
also covered by RGB, but not as evenly because of  its high optical flux limit.

\section{The Discovery of X-ray Bright FSRQ:  A Case in Point}

\begin{figure}
\epsfysize=9.0cm 
\hspace{1.5cm}\epsfbox{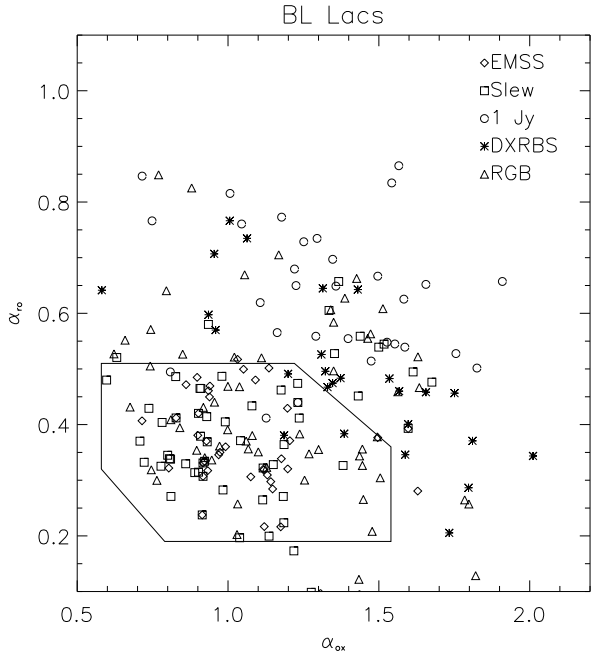}

\caption{The $(\alpha_{ox}, \alpha_{ro})$ plane for BL Lacs.  We have outlined 
the HBL "box" to guide the eye.  The 1 Jy, Slew and EMSS  contain very few
intermediate  objects. The results of the new DXRBS and RGB surveys show this
to be a selection effect.} 

\end{figure}

As shown above, the classical surveys are are far too small and shallow to
contain significant amounts of objects in all spectral shape classes.    Thus,
even when lumped together, they present a view that is  overwhelmingly colored
by biases.  For example, by comparing the broadband spectral shapes of the EMSS
and 1 Jy BL Lacs, and the S5 FSRQ, Sambruna et al. (1996) predicted that there
should be no FSRQ with $\alpha_{rx}\gsim 0.78$ - the historical dividing line
between HBL and LBL type objects.  This prediction was proven incorrect by the
findings of a newer, deeper radio-limited survey, the DXRBS (Perlman et al.
1998, Landt et al. 2000), which found that approximately 1 in 4 FSRQ had
$\alpha_{rx}\gsim 0.78$.  A subsequent cross-correlation of emission line
objects in the RGB (Laurent-Muehleisen et al. 1998) revealed an even higher
percentage of these objects ($\sim 40\%$, Padovani et al. in prep, Perlman
2000).

We will talk more about these objects (called ``HFSRQ'') below, but first a
historical comment is useful.  As is well known, the Slew and EMSS  both
included radio observations at 5 GHz in their identification process, but not
radio spectral indices.  The biases that imposed were not appreciated for
several years, even though both Stocke et al. (1991) and  Elvis et al. (1992)
noted the presence of radio-loud quasars in both surveys.  Only after the
results of DXRBS began to come out was this bias considered, and corrected. 
Cross-correlation of the Slew and EMSS object lists with lower-frequency radio
surveys reveal small, but significant samples of FSRQ in both (22 and 16
objects respectively; Perlman et al. 1999 and in prep.), of which
about 40\% are HFSRQ.  By comparison, the very high radio flux limit of the
1 Jy sample makes it almost completely insensitive to these objects (Perlman
et al. 1998).  Figure 2 summarizes these findings by comparing the FSRQ 
content of the new and classical surveys.

As was the case for HBL, HFSRQ cover a different region of $(L_x,L_r)$
parameter space than previously known FSRQ, and dominate increasingly at lower
luminosities.    As was pointed out by Urry \& Padovani (1995), the knee of the
FSRQ radio luminosity function sits at $L_r = 10^{33-33.5} {\rm erg s^{-1}
Hz^{-1}}$. A quick look at Figure 2 is enough to convince the reader that the 1
Jy survey had basically no sensitivity below that luminosity (in fact only 2 of
its objects have lower radio luminosities),  yet it is precisely at these
luminosities where the HFSRQ become increasingly common.

\section{X-ray Observations of HFSRQs}

\begin{figure} 
\epsfysize=9.0cm 
\hspace{1.5cm}\epsfbox{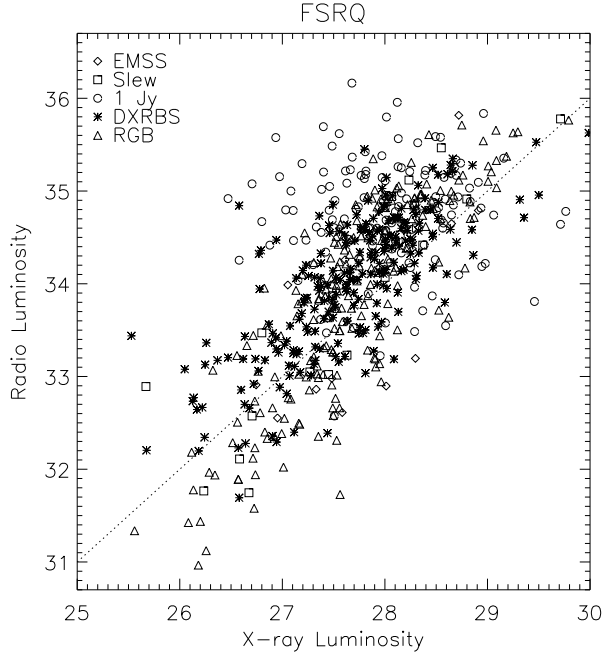}

\caption{The X-ray and Radio luminosities of FSRQ in the EMSS, Slew, 1 Jy,
DXRBS and RGB surveys.  The 1 Jy  contains only 14 FSRQ, ($\sim 5\%$) of FSRQ
to the right of the HBL/LBL  dividing line (dotted).  By comparison, 25\% of
DXRBS FSRQ, and 40\% of EMSS, Slew and RGB FSRQ, fall into this category.} 

\end{figure}

In order to understand the full spread of blazar properties, and develop
appropriate constraints upon their physics, it is imperative to investigate the
properties of HFSRQ in the same depth as has already been done for their BL Lac
cousins.  To start the process, we have observed four of the X-ray brightest
HFSRQ with SAX to analyze their 0.1-10 keV spectra.  In so doing, we aim to 
investigate the general trend noted in Perlman (2000) that DXRBS HFSRQ tend
to have somewhat steeper ROSAT spectra (as defined by their hardness ratios)
than their more radio-bright cousins. In Table 1, we give the results of these
observations (a full accounting of these results will be presented in a later
paper). 

\begin{table}
\caption{\normalsize{SAX Observations of HFSRQ - Results  }}
\begin{tabular}{lllll}
\hline
\hline
Name  & N$_H$ & $\alpha_{x}$ & $F_{1keV}$ & $\chi^2_r$/d.o.f.  \\
  & $10^{20}$ cm$^{-2}$ & & $\mu$Jy    \\ 
\hline
WGA J0546.6-6415 & 4.54 fix               & $0.72\pm0.08$ & $0.64\pm0.07$ &  1.07/41  \\
RGB J1629+401  & 0.852 fix              & $1.50\pm0.06$     & $0.66\pm0.05    $ & 0.90/28 \\
RGB J1722+243  & 4.95 fix         & $0.62\pm0.22$          & $0.16\pm0.05 $ & 0.63/18 \\
S5 2116+81:  \\
        .....29/2/98    & 7.41 fix               & $0.73\pm0.04$  & $2.71\pm0.17$   &  0.96/61\\
         ...12/10/98    & 7.41 fix               & $0.77\pm0.07$  & $2.15\pm0.19$   &  0.98/46  \\
   ...........sum       & 7.41 fix              & $0.73\pm0.04$  & $2.75\pm0.16$     & 1.04/62 \\
\hline
\hline
\multicolumn{5}{l}{\footnotesize{ Note: the errors are at $90\%$ conf. level for
 one  
parameter of interest. }} \\
\end{tabular}
\end{table}

As can be seen, we find flat spectra, more similar to LBL BL Lacs and
previously known FSRQ (Sambruna et al. 1996) for three of four objects.  Only
one object (RGBJ1629+401) has a steep, HBL-like spectrum.  This represents the
first systematic observation of objects with demonstrably HBL-like broadband
continua based upon their $\alpha_{ox}$ and $\alpha_{ro}$ values.  The ASCA 
observations of ``blue'' FSRQ by Sambruna et al. (2000) achieved an outwardy
similar result, with all objects having flat hard X-ray spectra.  However,
the Sambruna et al. result cannot really be considered as an observation of
HFSRQ, as their selection criteria was for steep ROSAT spectra and not 
$(\alpha_{ox}, \alpha_{ro})$ values indicative of a high frequency peak --
and indeed, some of the Sambruna et al. objects are {\it not} in the HBL
region of the $(\alpha_{ox}, \alpha_{ro})$ plane.

In this light, it is unclear how we should interpret the broadband spectra of
HFSRQ.  The most straightforward interpretation would be that in all cases
except for RGB J1629+401 we are seeing only inverse-Compton emission at  $>$0.1
keV.  However, extreme caution is required.   At the very least these objects
have two emission components (both synchrotron and emission line) which can
serve as  seeds for inverse-Compton emission - and therefore it is likely that
the  synchrotron to inverse-Compton ratio is considerably higher for HFSRQ than
it is for HBL BL Lacs. It is of course unclear where this additional emission 
might evidence itself, but according to the models of Sikora, Begelman \& Rees
(1994), inverse-Compton scattered emission-line photons should evidence
themselves at energies of order 1-100 keV.  Thus it is very likely that for
HFSRQ of a given peak frequency  the synchrotron to inverse-Compton ratio is
considerably higher than for an HBL of the same peak frequency (see
Georganopoulos 2000 for an interesting discussion of exactly this point).  Thus
it is still quite feasible that these objects have peak energies as high as
$\sim 10^{16}$ keV.  By contrast, for RGB J1629+401 the most consistent
explanation of its broadband spectrum would be that we are seeing synchrotron
emission all the way up to 10 keV.

\section{Discussion}

\begin{figure}
\epsfxsize=9.0cm 
\hspace{1.5cm}\epsfbox{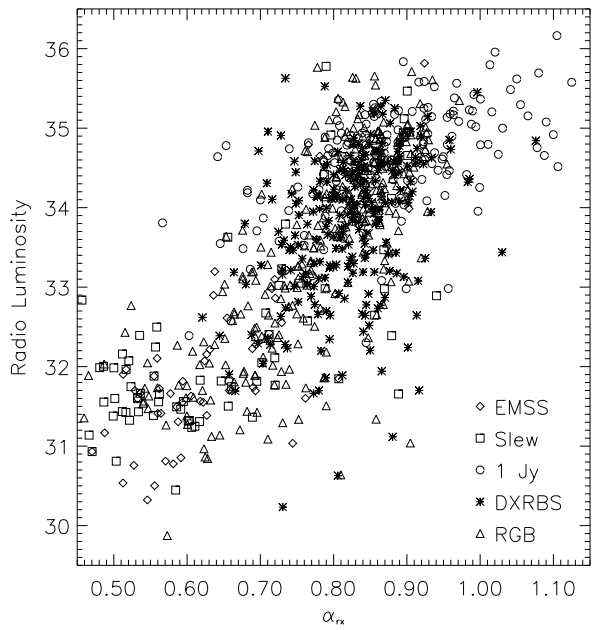}

\caption{The $(\alpha_{rx}, L_r)$ plane for blazars  including the 
EMSS, Slew, RGB, DXRBS and 1 Jy samples.  Note how much broader the correlation
becomes after all surveys are included.  While the correlation is still
$>99.99\%$ significant, the most prominent feature of this
graph is the prominent ``outlier'' regions at the upper left and lower
right hand corners, representing extreme objects.}

\end{figure}

It is now worthwhile to reassess the analysis of Fossati et al. (1998).  
Figure 3 shows the $(\alpha_{rx}, L_r)$ plane for blazars when all five blazar 
surveys are added to the diagram.  As discussed in Urry \& Padovani (1995), an
object's $\alpha_{rx}$ value is related to the location of its synchrotron
peak, with $\alpha_{rx} \approx 1$ representing  $\nu_{peak} \sim 10^{12}$ Hz,
and $\alpha_{rx} \approx 0.5$ representing  $\nu_{peak} \sim 10^{17}$ Hz.  
Thus, as explained by Fossati et al., the correlation between these two
parameters was meant to explore the relationship between total luminosity and
peak frequency. As can be seen, this correlation still persists, but when all
five samples are added, it grows quite a bit broader - in fact, most of the
horizontal extent in the graph is taken up by the ``outlier'' sections in the
lower left and upper right hand corners. These corners represent areas where
only one or two surveys had overwhelming sensitivity.  If one  only examines
the middle section of the graph, the correlation is still very significant, but
its physical meaningfulness is far less persuasive.  And in fact, when {\it
individual} surveys are examined no such correlation is present!  This is shown
convincingly by Figure 4, which is the same plot with only DXRBS objects
shown.  It is therefore impossible to tell whether the $(\alpha_{rx}, L_r)$
plot represents a physically important correlation, or is merely  reflective of
the parameter space sensitivity of each survey.

The comparison of Figures 3 and 4 illustrate quite well the dangers inherent in
combining surveys with very different biases and sensitivities, and attempting
to derive correlations. These biases will continue to color our perception of
blazar properties until  either radio and X-ray selected samples become large
and deep enough to include significant numbers of objects in every spectral
shape class.  DXRBS is the first radio-limited survey to accomplish this, but
it will take much deeper radio and optical surveys to accomplish this for an 
X-ray flux limited sample, even with ROSAT data (see Padovani 2000).  Thus to
fully understand the findings of surveys whose parameter space coverage lies in
the middle of the Padovani diagrams (e.g., REX, RGB), significant modeling will
be required.

Fossati (2000) has described a promising avenue for such modeling 
efforts.  However, it is instructive to note that the error bars on his models
are still quite large, probably because of the almost disjoint parameter space
coverage of the EMSS, Slew and 1 Jy surveys.  Once Fossati has included deeper
surveys such as DXRBS in his seed samples, it will be interesting to see whether
the resulting error bars are sufficiently small that the technique can be used
to predict the outcomes of future blazar surveys.  A similar comment applies to
the modelling efforts of Georganopoulos \& Marscher.

\begin{figure}
\epsfxsize=9.0cm 
\hspace{1.5cm}\epsfbox{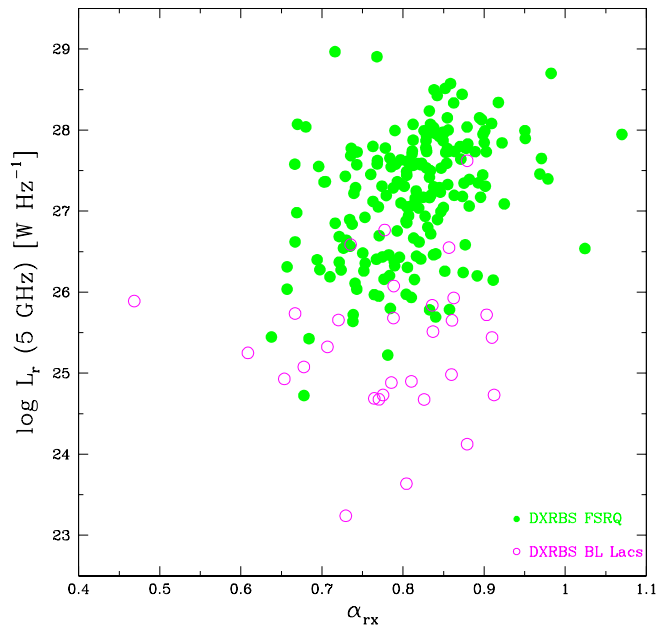}
\caption{The same plot as Figure 3, but with only DXRBS objects
shown.  As can be seen, the Fossati et al. correlation does not persist
when only single surveys are considered, making it quite conceivable that
the observed correlation reflects the parameter space sensitivity of each
survey rather than a physically relevant correlation.}
\end{figure}

\end{document}